\documentclass[nonacm,sigplan]{acmart}

\usepackage{tcolorbox, enumitem, afterpage}
\usepackage{multirow}

\AtBeginDocument{%
  \providecommand\BibTeX{{%
    \normalfont B\kern-0.5em{\scshape i\kern-0.25em b}\kern-0.8em\TeX}}}

\definecolor{ao(english)}{rgb}{0.0, 0.5, 0.0}
\begin{document}

\title[]{Diversity in Software Engineering Conferences and Journals}

\author{Aditya Shankar Narayanan}
\email{a8shanka@uwaterloo.ca}
\affiliation{%
  \institution{University of Waterloo}
  \city{Waterloo}
  \state{ON}
  \country{Canada}
  \postcode{N2L 3V8}
}

\author{Dheeraj Vagavolu}
\email{d2vagavo@uwaterloo.ca}
\affiliation{%
  \institution{University of Waterloo}
  \city{Waterloo}
  \state{ON}
  \country{Canada}
  \postcode{N2L 3V8}
}

\author{Nancy Day}
\email{nday@uwaterloo.ca}
\affiliation{%
  \institution{University of Waterloo}
  \city{Waterloo}
  \state{ON}
  \country{Canada}
  \postcode{N2L 3V8}
}

\author{Meiyappan Nagappan}
\email{mei.nagappan@uwaterloo.ca}
\affiliation{%
  \institution{University of Waterloo}
  \city{Waterloo}
  \state{ON}
  \country{Canada}
  \postcode{N2L 3V8}
}

\begin{abstract}
  Diversity with respect to ethnicity and gender has been studied in open-source and industrial settings for software development. Publication avenues such as academic conferences and journals contribute to the growing technology industry. However, there have been very few diversity-related studies conducted in the context of academia. In this paper, we study the ethnic, gender, and geographical diversity of the authors published in Software Engineering conferences and journals. We provide a systematic quantitative analysis of the diversity of publications and organizing and program committees of three top conferences and two top journals in Software Engineering, which indicates the existence of bias and entry barriers towards authors and committee members belonging to certain ethnicities, gender, and/or geographical locations in Software Engineering conferences and journal publications. For our study, we analyse publication (accepted authors) and committee data (Program and Organizing committee/ Journal Editorial Board) from the conferences ICSE, FSE, and ASE and the journals IEEE TSE and ACM TOSEM from 2010 to 2022. The analysis of the data shows that across participants and committee members, there are some communities that are consistently significantly lower in representation, for example, publications from countries in Africa, South America, and Oceania. However, a correlation study between the diversity of the committees and the participants did not yield any conclusive evidence. Furthermore, there is no conclusive evidence that papers with White authors or male authors were more likely to be cited. Finally, we see an improvement in the ethnic diversity of the authors over the years 2010-2022 but not in gender or geographical diversity. 
\end{abstract}




\maketitle

\section{Introduction}

The value of diversity in Software Engineering (SE) organizations has been well-documented in literature (e.g., ~\cite{se-diversity, se-productivity, se-productivity-2, se-productivity-3}). 
Østergaard et al.~\cite{se-productivity} and Tourani et al. ~\cite{se-productivity-2} show that gender diversity increases innovation and productivity, while Vasilescu et al. ~\cite{se-productivity-3} claim that perceived diversity of team members reduces turnover and conflict within software development teams. 
Nelson discusses the benefits of diversity for an organization, such as the higher earnings found at companies with more women in executive positions~\cite{Nelson2014}. Patents with diverse gender authorship are cited 26-42\% more than those with authors all of the same gender~\cite{Nelson2014}. 
Woolley and Malone find that the intelligence of a group is higher when there are more women in the group~\cite{Woolley2011}.
Diaz-Garcia et al. show that gender diversity in research and development teams leads to more radical innovations in companies in Spain~\cite{DiazGarcia2013}. Joecks et al. find that having more than 30\% female representation on supervisory boards of firms in Germany has a positive effect on a firm's performance~\cite{Joecks2013}. Nelson~\cite{Nelson2014} emphasize the value of collecting data on diversity. By collecting and analyzing data on diversity, the committee was able to identify areas where gender-based inequities existed and make recommendations for improving the status of women faculty at MIT~\cite{MIT1999}.

Literature shows a lack of diversity in various spaces of SE such as in the software development industry ~\cite{industry_diversity} and even Open Source Software (OSS) development ~\cite{oss_diversity, edi-oss}. Perez et al. ~\cite{se-diversity} perform a systematic literature review on the topic of perceived diversity, wherein they conclude that there is a need for more diversity-related studies in Software Engineering. Particularly, we find that very few diversity-related studies have been conducted in the context of Software Engineering academia. 

The Computing Research Association's annual Taulbee Study~\cite{taulbee2021} for universities with computing programs in the United States and Canada in 2021 reports that women received 23.3\% of the PhDs awarded in computer science for the institutions responding to the survey, which is an improvement over previous years. For hiring to tenure-track positions, 31.5\% of the positions went to women and overall 23.9\% of the faculty positions surveyed are held by women. With respect to ethnicity, 42.9\% of the new positions went to those of Asian ethnicity and 30\% went to those of White ethnicity, and overall, Asians (27.7\%) and Whites (54.2\%)  together hold 81.9\% of current faculty positions. Researchers are investigating ethnic disparity in the success rates for research funding and have found that scientists of Asian ethnicity are underrepresented~\cite{Chang2023}.
Much of the prior work studying diversity in academia, particularly the various domains in computer science, focuses on either providing mitigation strategies  (e.g.,~\cite{policy_diversity}), or on specific domains such as Artificial Intelligence~\cite{aiconf} or Systems Engineering~\cite{compsysauthors}. 

Within Software Engineering, governance in the most popular SE conferences such as the International Conference on Software Engineering (ICSE), ACM Joint European Software Engineering Conference and Symposium on the Foundations of Software Engineering (ESEC/FSE), the International Conference on Automated Software Engineering (ASE), Requirements Engineering (RE), and the International Conference on Evaluation and Assessment in Software Engineering (EASE) were found to have gender disparity, with female representation in visible roles such as general and program chairs as well as keynote speakers being less than 30\% on average ~\cite{governance_diversity}. 

In this paper, we shed light on the question of diversity in academic publishing for Software Engineering from 2010 to 2022. We provide a systematic quantitative analysis of the diversity of publications and organizing committees in prominent SE conferences and journals.

Academic conferences and journals in the various domains of computer science are a major avenue where researchers in the field can publish their work and obtain feedback from their peers. The works published and reviewed are precursors in the creation of technologies used in the software industry, to aid in or constitute a major part of the software development process. Therefore, the diversity found in the participants of academic conferences/journals could have an indirect impact on the diversity of the software industry. Ensuring that academic conferences/journals in Software Engineering are inclusive and that the authorship of the publications resulting from these conferences/journals is diverse is essential in developing strategies aimed at improving diversity and inclusion in Software Engineering spaces as a whole.

Our main contribution involves performing a systematic quantitative analysis of the authorship and governance metadata of three top-ranked academic conferences and two major journals for publishing Software Engineering research. We formulate research questions and derive insights based on the diversity attributes that are extracted from this data. Finally, we suggest mitigation strategies to address issues regarding the diversity of authorship in SE conferences and journals. The novelty of our work lies in the analysis of both authorship and governance committee diversity, our attempt to study the trends of their diversity over a period of 13 years, and in looking at both conferences and journals, particularly in SE.    

In this paper, we answer the following research questions (RQs):

\begin{enumerate}[label=\textbf{RQ\arabic*}]
    \item What is the diversity that exists in the authorship and governance of top SE conferences and journals?
    
    \item Does the diversity of the first author affect how often their work is cited by others? 

    \item Is the state of diversity improving in the authorship of SE conferences and journals?
    
    \item Is the diversity of the governance of SE conferences and journals correlated to the diversity of the authors of the works published at these venues?
\end{enumerate}

Section~\ref{sec:methodology} outlines our methodology to collect and process the data. Section~\ref{sec:results} presents the results for our research questions. Section~\ref{sec:discussion} discusses the implications of the results obtained, our insights and potential mitigation strategies to address the issues pertaining to diversity and inclusion in SE conferences and journals. Section~\ref{sec:threats} explains the various threats to the validity of our findings as well as how we address these threats. Section~\ref{sec:relatedwork} provides a brief overview of the existing literature related to the study of diversity in SE and also provides a comparison between the related studies and our paper.  Finally, Section~\ref{sec:conclusion} summarizes our methodology and findings.

\section{Methodology}
\label{sec:methodology}

\begin{figure*}[!ht]
  \centering
  \includegraphics[scale=0.55]{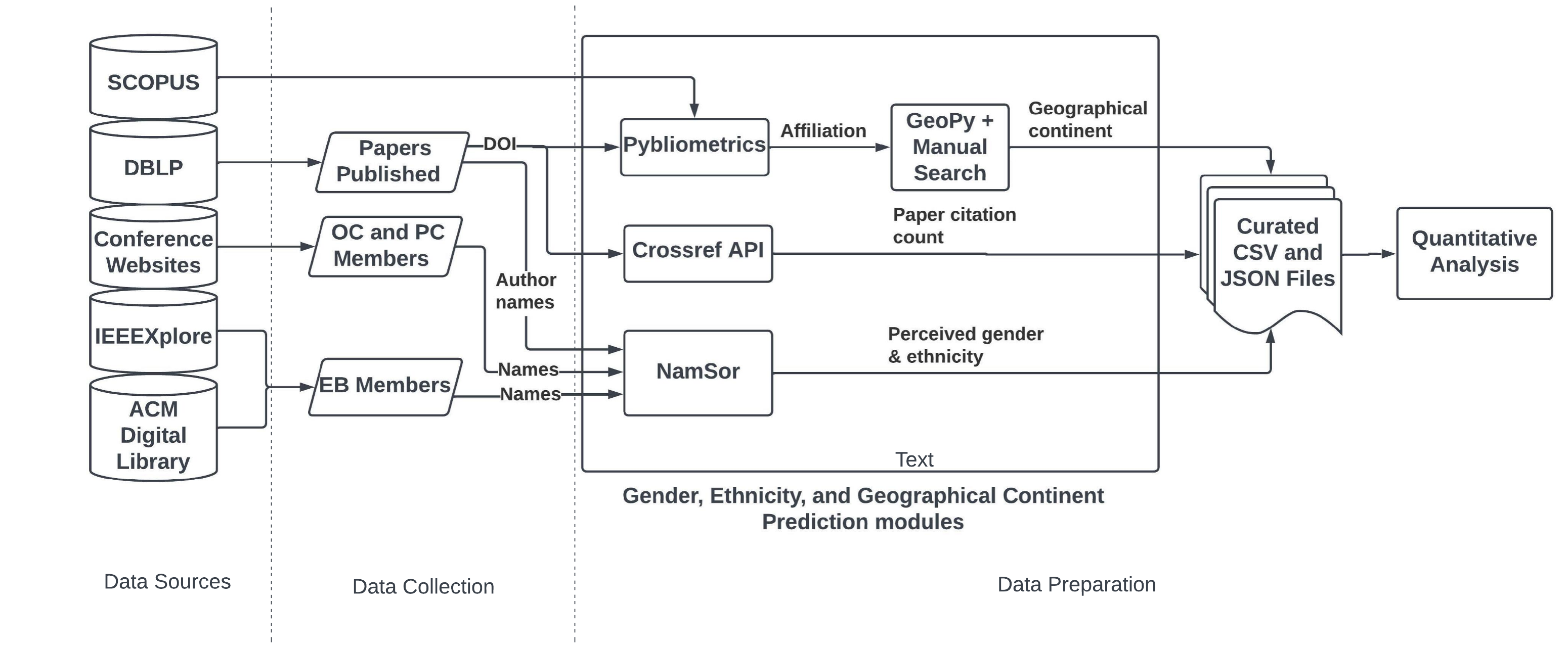}
  \caption{Methodology for evaluating diversity in SE conferences/journals.}
  \label{fig:pipeline}
\end{figure*}

In this section, we discuss the sources of the data used for this study and our general approach used in the collection and processing of this data. To answer our research questions, we require data about the papers published at major SE academic venues. We select the usually considered top three general SE academic conferences - the International Conference on Software Engineering (ICSE, Main track)~\cite{icse}, ACM Joint European Software Engineering Conference and Symposium on the Foundations of Software Engineering (ESEC/FSE -- shortened to FSE for brevity in this paper)~\cite{esecfse2023}, and International Conference on Automated Software Engineering (ASE) ~\cite{ase2023}
(for conference rankings see~\cite{ConferenceRank2023}). We also include two highly ranked SE journals that cover all topics in SE - the IEEE Transactions on Software Engineering (IEEE TSE) ~\cite{tse}, and ACM Transactions on Software Engineering and Methodology (ACM TOSEM) ~\cite{tosem} - in our analysis (see journal rankings at~\cite{JournalRank2023}). Figure~\ref{fig:pipeline} illustrates our methodology for the collection, processing, and analysis of the data, which is explained in the following subsections.

\subsection{Data Sources}
\label{source}

From the DBLP computer science on-line bibliography~\cite{DBLP}, we obtain the list of paper titles, the corresponding DOIs, and the list of author names among other publishing metadata for all the papers published at the three conferences and two journals over the period of 2010-2022.  For data about the governance of these venues, we find the list of program committee (PC) and organizing committee (OC) members for each year and conference from the official conference websites (for the more recent years) and the front matter of the published transactions (for the older years). This list provides us with the names and affiliated institutions of the committee members. We also get the list of journal editorial board (EB) members (and their affiliations) of ACM TOSEM over the years from the front matter of the transactions published on ACM Digital Library ~\cite{acmDigitalLibrary}. For IEEE TSE, we are able to find the past EB data only for the years 2010 to 2013 from the front matter of the Transactions published on IEEE Xplore ~\cite{ieeeIEEEXplore}. The data on the IEEE TSE EB for 2014-2022 is not publicly available. Thus, the analysis for the governance of IEEE TSE is limited by the lack of availability of this data.  

\subsection{Data Collection}

We programmatically query DBLP  by means of dblp-retriever \cite{dblp-retriever} to retrieve the publishing information. The name of the conference, the year, and the corresponding DBLP conference identifier are provided as inputs to the tool. The dblp-retriever returns the following metadata of each paper: venue, year, identifier, heading, title, authors, pages, length, and electronic edition (DOI).  In total, we obtain 35782 papers across all the conferences and journals.

For the conferences, we manually extract the names and affiliations of each member of the PC and OC from the official conference websites, going as far back as 2016. For 2010-2015, we manually obtain this information from the front matter of the published proceedings on IEEE Xplore and ACM Digital Library. Similarly, we get the EB members for the two journals from the front matter of the published transactions across all the issues in the years 2010-2022 from IEEE Xplore and ACM Digital Library (with the exception of IEEE TSE 2014-2022, as noted in Section~\ref{source}). To remain consistent, we combine the list of editors for all the issues within the same volume (which is referenced by the year of publication). Therefore, there is an underlying assumption that all the editors that come under the same year were present on the EB of all the issues in that year.

\begin{figure*}[t]
  \centering
  \includegraphics[scale=0.3]{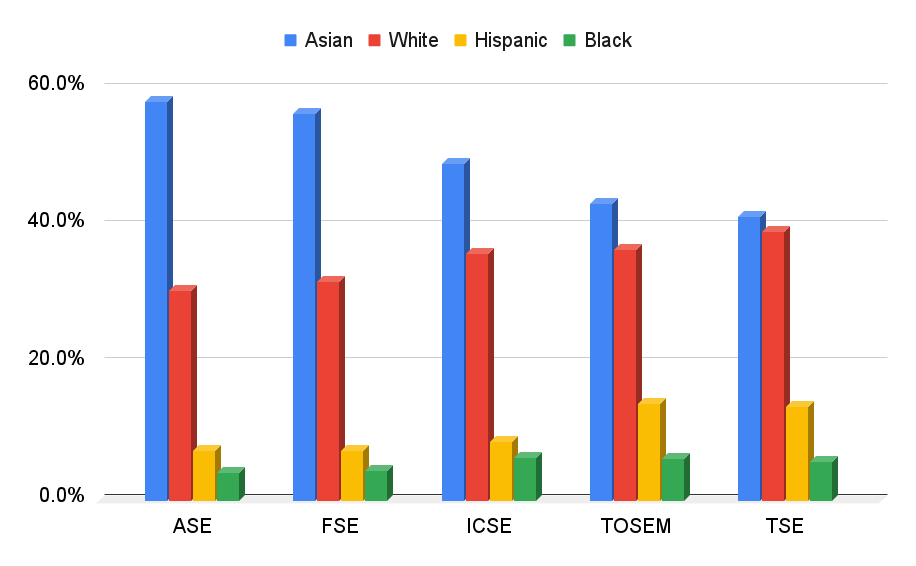}
  \caption{Ethnic diversity of authors.}
  \label{fig:rq1_eth}
  \includegraphics[scale=0.35]{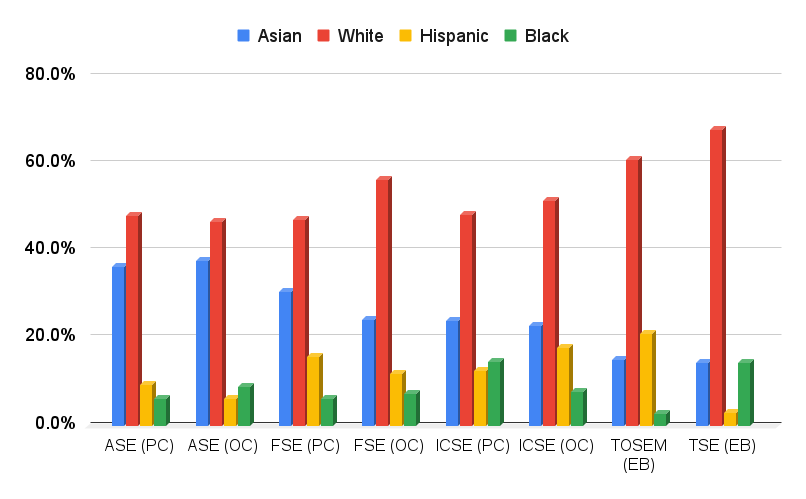}
  \caption{Ethnic diversity of governing members.}
  \label{fig:rq1_eth_gov}
\end{figure*}

\subsection{Data Preparation}

We use a Python module called Pybliometrics ~\cite{pybliometrics} that queries the SCOPUS database and returns the affiliated institution as well as the country of each author of a given paper's DOI (at the time the paper was published). SCOPUS ~\cite{scopus} is a curated database that indexes scholarly articles from a variety of academic publications. The database consists of metadata on the published articles such as abstracts, citation count, list of authors and their affiliated institutions (at the time the article was published) and the corresponding country. SCOPUS also provides APIs for querying the database either with the title or with the DOI of the article. 

To find the geographical location of the published authors and the committee members, we use the country information from the affiliated institution of each person. We get the affiliated institutions and the corresponding country for the authors of the papers (at the time the paper was published) from SCOPUS. However, for the members of the PC, OC, and EB, we get the affiliated institutions along with the names from the corresponding data sources (as mentioned in Section~\ref{source}). From this data, we extract the country associated with the affiliated institutions of the committee members using a Python module called GeoPy~\cite{geopy}.  For about 150 institutions, we manually determine the geographical location of an institution. We then divide the geographical location data into the six continents (Asia, Europe, Africa, North America, South America, and Oceania) of each location.

To answer RQ2, we require the number of times each published paper has been cited to date. We use Crossref's Python API ~\cite{pypiCrossrefapi} that takes the DOI of a paper as input, queries the Crossref database~\cite{crossref}, and returns the number of times the paper has been referenced by other published work. 


Next, we determine the diversity of the authors and committee members based on their names. We use NamSor~\cite{namsor} to determine the perceived gender and ethnicity of every author published in any of the conferences and journals, along with every member for the PC, OC and EB. NamSor provides a web interface that can take a .csv or .xlsx file containing a list of names separated into given name and surname as input and predicts the gender using the "Genderize Name from first and last Name" feature (two classes: male and female) and the ethnicity using the "Name US Race" feature  (four classes: White, Asian, Hispanic, and Black) for each name.

The result of our data preparation process is a set of curated CSV and JSON files containing information on the published papers, their authors, the gender, ethnicity and geographical location of these authors, and the same information for members of PC, OC, and EBs for the three SE conferences and two SE journals included in our study.

\section{Results}
\label{sec:results}

In this section, we present the results for our four research questions regarding the diversity in gender (male or female), ethnicity (White, Asian, Hispanic, or Black), and geographical location (continent of the location of a person's affiliated institution) found in SE conferences and journals.

\subsection{RQ1}
\begin{tcolorbox}
[width=\linewidth, sharp corners=all, colback=white!95!black]
\textbf{RQ1:} What is the diversity  that exists in the authorship and governance of top SE conferences and journals?
\end{tcolorbox}

Lack of diversity is a prominent challenge observed across several technical fields such as the software development industry and the OSS community \cite{oss_diversity, industry_diversity}. Similar challenges have been faced by the academic research communities specifically in engineering domains \cite{governance_diversity}. Freire et al. shed light on the lack of diversity present in Artificial Intelligence conferences based on gender, geographical location, and the involvement of industry \cite{aiconf}. Bano and Zowghi study the diversity in the governance of SE conferences by measuring the diversity in the visible roles of academic conferences such as the program chairs and the committee members~\cite{governance_diversity}.

To our knowledge, very few works focus on the diversity of those who publish in academia, and no work has aimed to measure the diversity specifically in SE conferences/journals. In RQ1, our goal is to fill this gap and report the diversity measures among authors in some of the major SE conferences and journals. 

\subsubsection{Approach}

We approach RQ1 by visualizing the curated raw data obtained for the three conferences and two journals. We combine all the authors who have published at all the selected venues over all the years and visualize the percentages of each of the four ethnicity classifications, two gender classifications, and six geographical location classifications in the form of bar graphs. We follow a similar process for the governing committees (PC/OC/EB) of these venues.

\begin{figure*}[!ht]
  \includegraphics[width=0.45\textwidth]{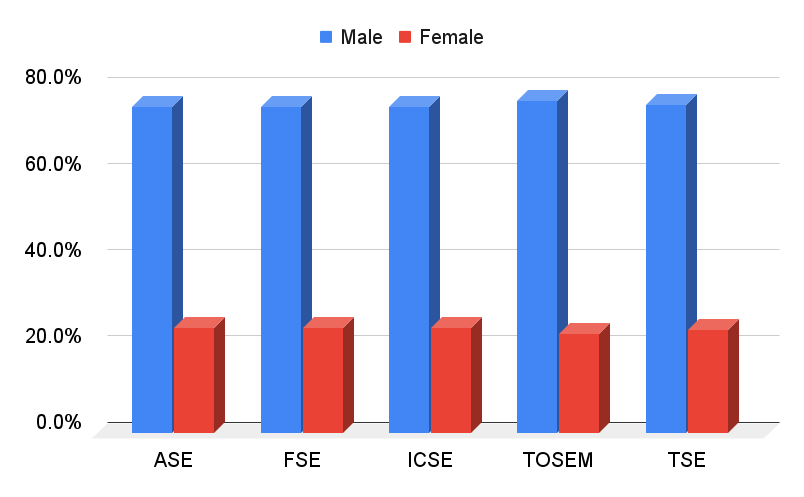}
  \caption{Gender diversity of authors.}
  \label{fig:rq1_gen}
  \includegraphics[width=0.45\textwidth]{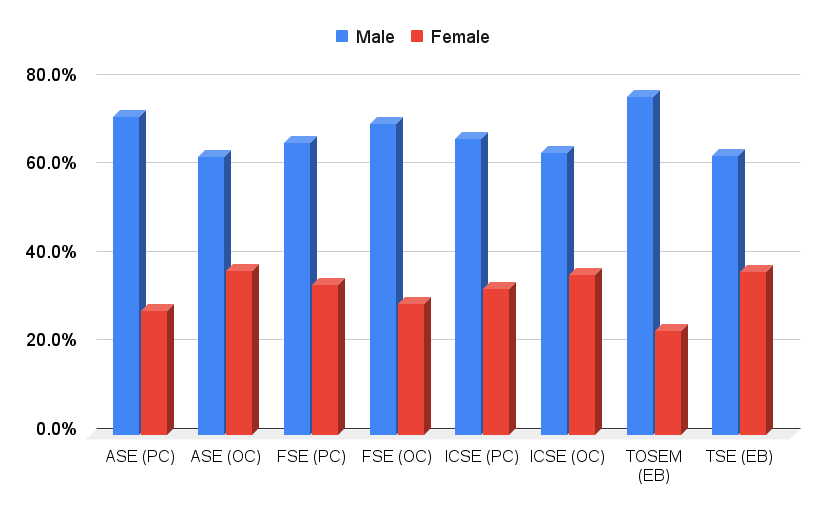}
  \caption{Gender diversity of governing members.}
  \label{fig:rq1_gen_gov}
\end{figure*}

\begin{figure*}[!ht]
  \centering
  \includegraphics[width=0.6\textwidth]{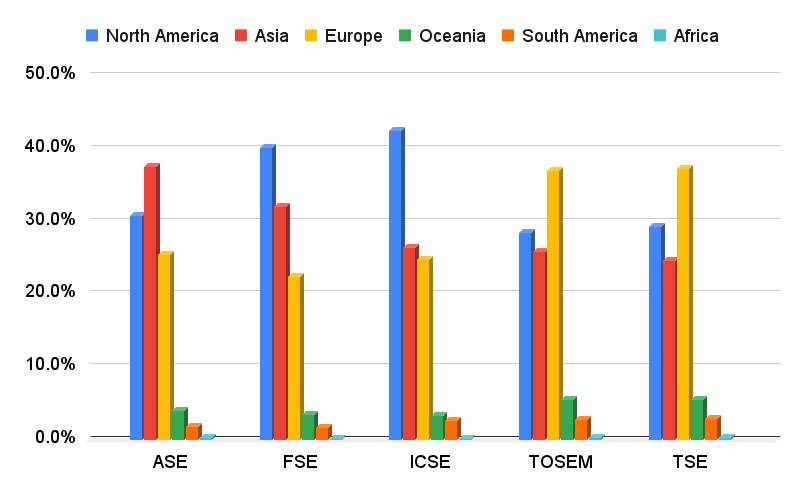}
  \caption{Geographical diversity of authors.}
  \label{fig:rq1_geo}
  
  \centering
  \includegraphics[width=0.7\textwidth]{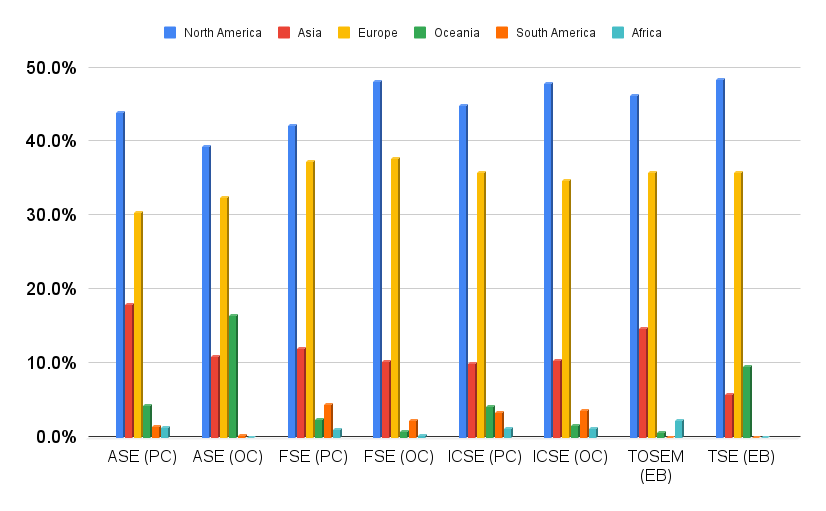}
  \caption{Geographical diversity of governing members.}
  \label{fig:rq1_geo_gov}
\end{figure*}

\subsubsection{Findings}

\textbf{Ethnic diversity.} The mean of the total number of papers accepted across the venues, regardless of the ethnicity of the authors, is 5054 across all the years (2010-2022). ICSE has the highest number of accepted papers with 7182, and the journal ACM TOSEM has the lowest with 1692 accepted papers across all the years. When comparing the presence of authors of a particular ethnicity at a venue, we consider the variable sample size and use relative frequency to observe the distribution of authors across the venues. From Figure \ref{fig:rq1_eth}, the highest percentage of any ethnicity is observed at ASE for Asian authors, comprising of 57\% of the total accepted papers across all the years. In contrast, Black authors constitute 4\% of total accepted papers at ASE, which is the lowest percentage observed. A similar distribution, with the highest share occupied by the Asian authors followed by White, Hispanic and Black authors, can be seen across all the five venues. However, the journals have a lower standard deviations across ethnicities when compared to the conferences.

Figure \ref{fig:rq1_eth_gov} shows the relative ethnicity of the governance committee members at the conferences and journals. Members with White ethnicity have the highest proportions, ranging from 48\% to 56.3\% in the conferences. For journals, the percentage of EB members with White ethnicity are 61\% at ACM TOSEM and 79\% at IEEE TSE. For PC and OC members at the conferences, the second highest proportion is occupied by Asian members, ranging from 22\% at ICSE OC to 37.8\% at ASE OC. At ACM TOSEM the percentage of Hispanic EB members (21\%) is greater than the percentage of Asian EB members (15\%).

\textbf{Gender diversity.} The gender distribution of authors across all the venues over all the years combined is uniform, with a maximum difference of about 4\% between the venues, as can be inferred from Figure~\ref{fig:rq1_gen}. The proportion of male authors is 74\%-76\%. Figure~\ref{fig:rq1_gen_gov} shows the gender diversity of the governing members. At the conferences, the proportion of male governance committee members ranges from 62\% for the ASE OC to 71\% for ASE PC.
The EB of ACM TOSEM shows a similar distribution as the gender distribution of the authors of all the venues, with 24\% female and 76\% male members. The distribution is comparatively better for IEEE TSE with 37\% female and 63\% male members. Overall, the gender diversity of governing committee members is slightly better than that of the authors at the conferences.

\textbf{Geographical diversity.} As shown in Figure~\ref{fig:rq1_geo} (authors) and Figure~\ref{fig:rq1_geo_gov} (governance), the geographical location of the authors at the venues can be clearly split into two categories. The top three continents - Europe, Asia and North America - together dominate both publications and governance, whereas Oceania, South America, and Africa have very low representation across all the venues. At ASE, the highest proportion of authors is occupied by Asia followed by North America. In the case of FSE and ICSE, North America accounts for the highest number of authors. ACM TOSEM and IEEE TSE have the highest number of authors affiliated with Europe, with a share of around 38\% for both. The lowest percentage of authors is from Africa, consisting of 1\% to 2\% across the venues.

In the case of the governing committee members, the highest proportion of the members are consistently affiliated with North American institutions, with an average of 44\% for the conferences and 46\% for the journals. The second highest proportion is held by the members affiliated with European institutions, with an average of 34\% for conferences and 36\% for the journals. Members who are affiliated with Asian institutions range from 10\% to 18\% at the conferences, with the highest being 18\% for ASE PC members. For the journals IEEE TSE and ACM TOSEM, the percentage of members from Asia is around 5\% and 14\% respectively. Similar to the authors, the proportion of African members in the governing committees range from 1\% to 2\%, and South America from 1\% to 4\%. There are no committee members with affiliations from South America on the EB of the journals.

\textbf{Summary.} When looking at the authors of the venues, the highest percentage of the population belongs to the Asian ethnicity followed by the White ethnicity. However, this trend is reversed when looking at the governing members. From Figure \ref{fig:rq1_eth}, the standard deviation of the ethnicities is higher in the conferences as compared to the journals for authors, but the reverse is true when looking at the distribution of governing committee members from Figure~\ref{fig:rq1_eth_gov}. We also find that a lack of gender-based diversity is consistent and very prominent at all venues among authors as well as PC/OC/EB members. For the geographical diversity, we see that countries from Oceania, South America and Africa have consistently low authorship and governing committee members across all the venues. Interestingly, the proportion occupied by authors affiliated with Asia is much higher on average when compared to the governing committee members affiliated with Asia across all the conferences.

\subsection{RQ2} 

\begin{tcolorbox}
[width=\linewidth, sharp corners=all, colback=white!95!black]
\textbf{RQ2:} Does the diversity of the first author affect how often their
work is cited by others?
\end{tcolorbox}

Our second research question aims to investigate the relationship between the gender, ethnicity, and geographical location of the first author of papers accepted by the conferences/journals and their respective citation counts (number of times their paper has been referenced by another work to date). We seek to determine if there are signs of any biases based on gender, ethnicity, and geographical location that researchers may have when citing the published works.

Studies have shown that conscious or unconscious biases can hinder the decisions made by people at various organizations \cite{social_factors_industry}. Ford et al. performed an eye tracking study to demonstrate that social aspects also play a vital role in OSS communities, particularly in the review of pull requests \cite{social_factors_oss}. If similar biases affect the decisions made by authors while finding and citing academic documents, it can poorly affect the visibility and recognition of researchers belonging to some communities.
We hypothesize that there is a statistical difference in the mean citation count of research papers categorized based on the ethnicity/gender/geographical location of the first author. 

\subsubsection{Approach}

Using the ANOVA test ~\cite{anova-test}, we first find out for which venues and years was there a significant difference in the means of the citations, depending on the category of diversity. We display the venue, they year, the first rank as determined by the Scott-Knott test, and the category of diversity for which we find a significant statistical difference in the mean citation count of papers in Table 1. We then use Scott-Knott clustering algorithm ~\cite{scott-knott-test} to see which particular subcategories were statistically different, and to find any patterns amongst them.  

\subsubsection{Findings}

\begin{table}[]
\caption{Venue, categories and the first rank determined by the Scott-Knott test in the years where there is a statistical difference in the mean citation count of the categories.}

\label{tab:citation_anova}
\begin{tabular}{|p{0.75in}|c|p{1in}|c|}
\hline
 \textbf{Venue} & \textbf{Year} & \textbf{Category} & \textbf{First Rank}\\ \hline 
 \multirow{4}{*}{ASE} & 2015 & Ethnicity & Asian \\
 & 2016 & Ethnicity & Black \\
 & 2018 & Ethnicity & Asian \\
 & 2020 & Ethnicity & Asian \\
 \hline
 \multirow{2}{*}{FSE} & 2013 & Geographical Location & North America \\
& 2014 & Geographical Location & North America\\
\hline
\multirow{1}{*}{ICSE} & 2012 & Ethnicity & Asian, Black \\
\hline
\multirow{3}{*}{ACM TOSEM} & 2011 & Geographical Location & North America \\
& 2013 & Geographical Location & North America \\
& 2019 & Gender & Female\\
\hline
\multirow{2}{*}{IEEE TSE} & 2012 & Gender & Female \\
& 2021 & Ethnicity & Asian  \\
\hline

\end{tabular}
\end{table}

For three out of four years that have a significant statistical difference of mean citation counts at ASE (2015, 2018, and 2020) from table ~\ref{tab:citation_anova}, there is a greater likelihood of having a higher citation count if the ethnicity of the first author is Asian. In 2016, the Asian ethnicity is ranked second to the Black ethnicity for citation count. The other statistically significant entries in the ethnicity category are ICSE in 2012 and IEEE TSE in 2021, both of which show first rank for Asian ethnicity. ICSE in 2012 is the only other venue which also shows Black ethnicity as first rank, along with ASE 2016. The ranking of the other ethnicities at all the other venues and years do not show any patterns, so no conclusions are drawn about them.

For FSE, the ANOVA test suggests that the citation count is affected by the geographical location of the authors. The Scott-Knott test ranks North America, Europe and Asia as the top three continents for both 2013 and 2014. 

For ACM TOSEM 2019 and IEEE TSE 2012, the Scott-Knott test shows that the chance of having a higher citation count is more likely if the gender of the first author is female. However, for the rest of the years there is no statistical significance between the gender and the citation count.


In summary, based on the results of the ANOVA and Scott-Knott tests, we are unable to show any conclusive evidence of systematic biases, based on gender, ethnicity, and geographical locations of first authors of papers, that researchers may have when citing the published works from these venues.

\subsection{RQ3}

\begin{tcolorbox}
[width=\linewidth, sharp corners=all, colback=white!95!black]
\textbf{RQ3:} Is the state of diversity improving in the
authorship of SE conferences and journals?
\end{tcolorbox}

Our third research question examines whether the efforts to improve diversity and inclusivity of authorship in these top conferences and journals have been effective. Since, the changes between the committee members each year is limited, particularly in the case of EB members of journals, and we lack data for IEEE TSE EB from the year 2014 onwards, we exclude the governing committees from the analysis done in RQ3. 

Several works propose and outline policies aimed at improving the lack of diversity in conferences and journals. For example, Serebrenik et al. outline several initiatives and efforts undertaken by ACM SIGSOFT to address the issues surrounding the lack of diversity~\cite{policy_diversity}. In RQ3, we aim to understand the usefulness and the impact of such initiatives by measuring the evolution of diversity of the authors over the past thirteen years. 

\subsubsection{Approach}
To study the evolution of diversity of authors in SE conferences and journals, we generate line graphs using the Blau diversity index ~\cite{blau1977inequality} for each conference/journal and category of diversity. The graphs show the measured Blau diversity index for ethnicity, gender, and geographical location of the authors plotted against the years 2010-2022 as dotted lines. The Blau diversity index is a value between 0 and 1, where 0 denotes a completely homogeneous population with no diversity and 1 denotes a uniformly distributed population across the different diversity classes. We find the best-fit line across the years (denoted as a solid line in Figures ~\ref{fig:rq3_eth}, ~\ref{fig:rq3_gen}, ~\ref{fig:rq3_geo}), and compare the slopes to see if there is an overall increasing or an overall decreasing trend. 

\subsubsection{Findings}

\begin{figure}[!ht]
  \centering
  \includegraphics[scale=0.37]{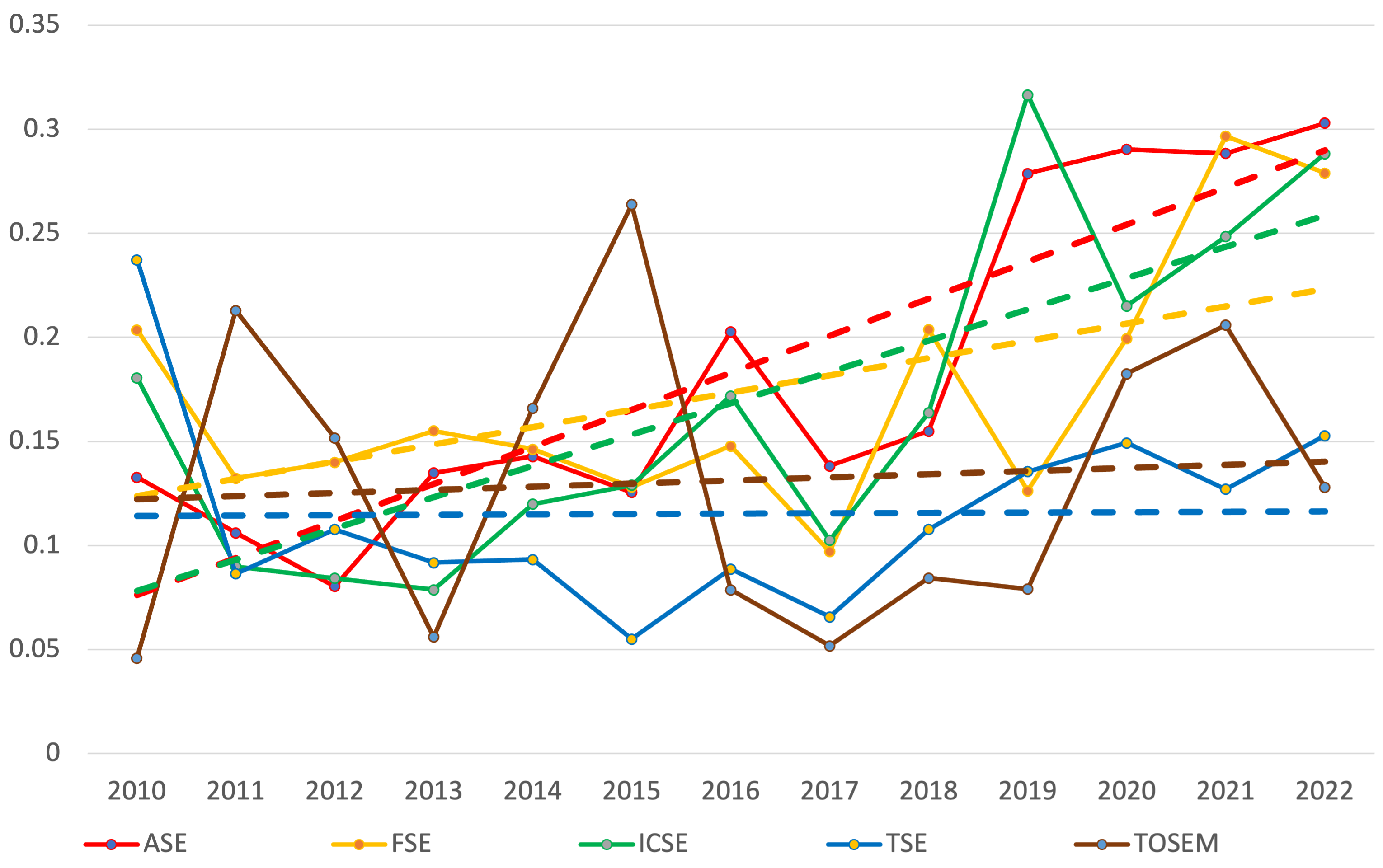}
  \caption{Ethnic diversity (Blau index) of authors from 2010 to 2022.}
  \label{fig:rq3_eth}
  \includegraphics[scale=0.37]{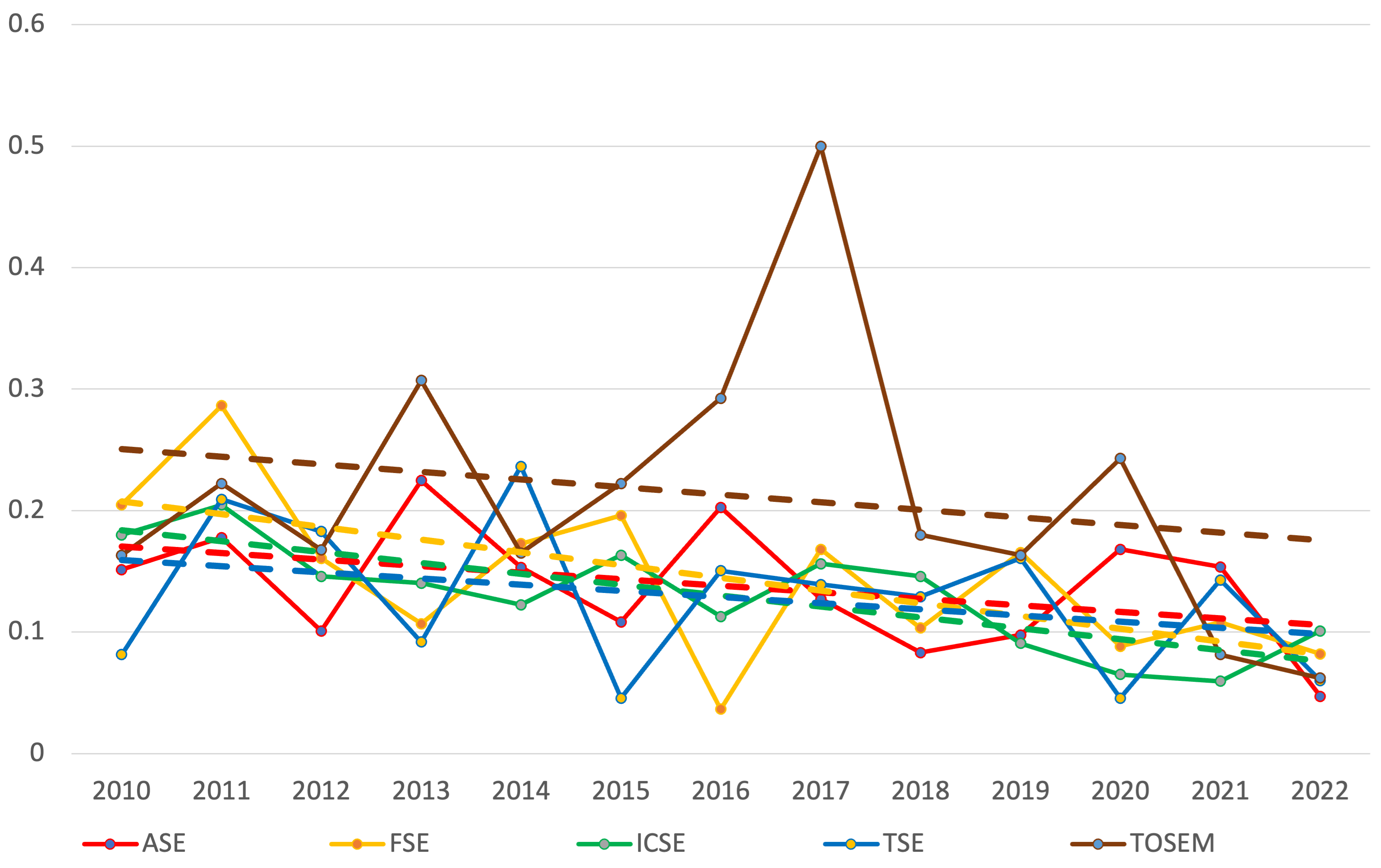}
  \caption{Gender diversity (Blau index) of authors from 2010 to 2022.}
  \label{fig:rq3_gen}
  \includegraphics[scale=0.37]{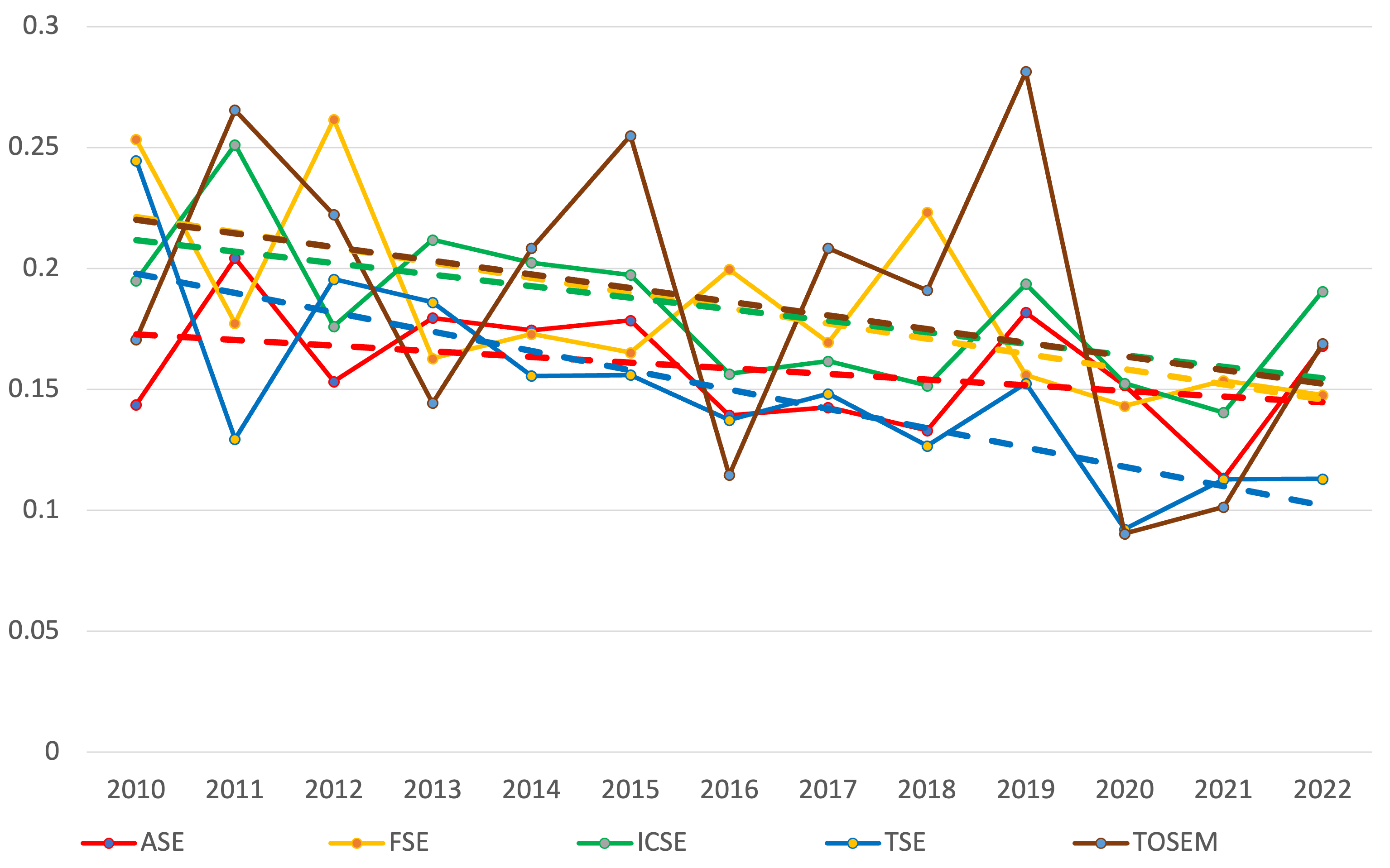}
  \caption{Geographical diversity (Blau index) of authors from 2010 to 2022.}
  \label{fig:rq3_geo}
\end{figure}

\noindent 

Figures~\ref{fig:rq3_eth},~\ref{fig:rq3_gen} and~\ref{fig:rq3_geo} show the Blau indices for the ethnicity, gender, and geographical location of the authors respectively, for the venues across all the years. The solid line indicates the line graph plot for the Blau indices for each conference, whereas the dashed line indicates the trend of the values. The Blau indices for the ethnicity of authors across all venues ranges from 0.04 to 0.3. Similarly, for gender and geographical location, the respective ranges are from 0.03 to 0.5 and from 0.1 to 0.28 respectively. For the ethnicity of authors, we see that the Blau indices show a general upward trend for the conferences ASE, ICSE and FSE, in that order. However, the journals ACM TOSEM and IEEE TSE have a relatively flat slope, indicating a neutral trend for the diversity in the ethnicities. For the gender and geographical location of authors, the Blau indices show a downward slope across all the venues. 

The standard deviation of ACM TOSEM is higher when compared to the other venues, for all the diversity categories. Hence, even though ACM TOSEM has some of the highest Blau indices across all the three categories (ethnicity, gender and geographical location), the slope is neutral for ethnicity and negative for gender and geographical location. 

In summary, we observe that only the ethnic diversity of the authors published at the SE conferences has improved over the past decade and not the gender and geographical diversity.

\subsection{RQ4}

\begin{tcolorbox}
[width=\linewidth, sharp corners=all, colback=white!95!black]
\textbf{RQ4:} Is the diversity of the governance of SE conferences and
journals correlated to the diversity of the authors of the
works published at these venues?
\end{tcolorbox}

To answer RQ4, we conduct a Pearson correlation test on the ethnicity, gender, and geographical diversity of authors and governance committee members. Bano and Zowghi demonstrate the lack of diversity in the visible roles in SE conferences and how they affect the hierarchy of roles \cite{governance_diversity}. We hypothesize that the lack of diversity in the governance of SE conferences directly affects the lack of diversity in the participants/authors. Hence, in this RQ4, we aim to confirm this hypothesis based on the correlation of diversity measures amongst the committees and the authors of major SE conferences.

\subsubsection{Approach}

In this RQ, we test for a correlation between the Blau diversity indices of the authors published in the conferences/journals and the Blau diversity indices of the corresponding PC/OC/EB members for the ethnicity, gender and geographical diversity across all the years combined using Pearson correlation test. To measure the strength of the correlation, we use the commonly accepted thresholds for interpreting the strength of a Pearson coefficient (r), where 0.40 <= r <= 0.59 is moderate, 0.60 <= r <= 0.79 is strong and 0.80 <= r is very strong. A positive coefficient indicates a direct correlation between the Blau indices of the authors and the committee members, whereas a negative coefficient indicates an inverse correlation between them.

\begin{table}[]

\caption{Pearson coefficient for ethnicity of authors and committee members across all conferences and journals.}
\label{tab:eth_pearson}
\begin{tabular}{|l|c|c|c|p{0.5in}|p{0.5in}|}
\hline
 & \textbf{ASE} & \textbf{FSE} & \textbf{ICSE} & \textbf{ACM TOSEM} & \textbf{IEEE TSE} \\ \hline
\textbf{PC} & \textcolor{red}{-0.482} & -0.332 & -0.139 & - & -  \\ \hline
\textbf{OC} & -0.177 & -0.286 & \textcolor{red}{-0.463} & - & -  \\ \hline
\textbf{Editors} & - & - & - & -0.353 & -0.285 \\ \hline

\end{tabular}

\vspace*{0.3cm}

\caption{Pearson coefficient for gender of authors and committee members across all conferences and journals.}

\label{tab:gen_pearson}
\begin{tabular}{|l|c|c|c|p{0.5in}|p{0.5in}|}
\hline
 & \textbf{ASE} & \textbf{FSE} & \textbf{ICSE} & \textbf{ACM TOSEM} & \textbf{IEEE TSE} \\ \hline
\textbf{PC} & \textcolor{red}{0.466} & 0.227 & 0.309 & - & - \\ \hline
\textbf{OC} & 0.019 & \textcolor{red}{0.411} & 0.091 & - & - \\ \hline
\textbf{Editors} & - & - & - & \textcolor{red}{-0.581} & \textcolor{ao(english)}{-0.881} \\ \hline
\end{tabular}

\vspace*{0.3cm}

\caption{Pearson coefficient for geographical location of authors and committee members across all conferences and journals.}

\label{tab:geo_pearson}
\begin{tabular}{|l|c|c|c|p{0.5in}|p{0.5in}|}
\hline
 & \textbf{ASE} & \textbf{FSE} & \textbf{ICSE} & \textbf{ACM TOSEM} & \textbf{IEEE TSE} \\ \hline
\textbf{PC} & \textcolor{blue}{0.603} & \textcolor{red}{0.445} & 0.338 & - & - \\ \hline
\textbf{OC} & -0.136 & \textcolor{blue}{0.751} & 0.035 & - & - \\ \hline
\textbf{Editors} & - & - & - & 0.204 & -0.255 \\ \hline
\end{tabular}

\vspace*{0.3cm}

\begin{tcolorbox}
[width=\linewidth, sharp corners=all, colback=white!99!black, valign=center, halign=center]
\textbf{\textcolor{red}{Red}: Moderate, \textcolor{blue}{Blue}: Strong, \textcolor{ao(english)}{Green}: Very Strong}
\textit{(Color Codes for Pearson Coefficient)}
\end{tcolorbox}

\vspace*{-0.8cm}
\end{table}

\noindent

\subsubsection{Findings}
Tables~\ref{tab:eth_pearson},~\ref{tab:gen_pearson}, and~\ref{tab:geo_pearson} show the Pearson coefficients for the diversity index for each of the PC/OC/EB members against the authorship, for IEEE TSE across the years 2010 to 2013 and the years 2010 to 2022 for the other venues. 

From Table \ref{tab:eth_pearson}, we observe a moderate negative correlation between the ethnic diversity of the authors and the PC members at ASE (-0.482), and a moderate negative correlation between the authors and OC members at ICSE (-0.463). The diversity of authors at the other venues did not exhibit any remarkable correlation with their respective OC/PC/EB members. 

From Table \ref{tab:gen_pearson}, we observe a moderately positive correlation between the gender diversity of the authors and PC members for ASE (0.466) as well as the authors and OC members for FSE (0.411).
For the journals, the Pearson coefficients for the gender diversity of the authors and EB members are moderately negative for ACM TOSEM (-0.581) and very strongly negative for IEEE TSE (-0.881). 
Due to a lack of data on the EB members of IEEE TSE from the year 2014 onwards, the Pearson coefficient only reflects the correlation of the diversity of the authors and the EB members for the years 2010-2013 for IEEE TSE, limiting the significance of the strong negative correlation (-0.881). 

From Table \ref{tab:geo_pearson}, we find a strong positive correlation between the geographical diversity of the authors and the PC members at ASE (0.603). For the geographical diversity at FSE, a moderate correlation is observed between the authors and the PC members (0.445), whereas a strong correlation is found between the authors and the OC members (0.751).

Overall, we find no significant correlation between the ethnicity and geographical diversity of the authors and the EB members for either of the journals. In contrast, we find a strong positive correlation in the diversity of the authors and the PC members of ASE as well as the authors and the OC members of FSE for geographical diversity. No other patterns were apparent for the correlation between the diversity of the authors and PC/OC/EB members of the other conferences and journals. Furthermore, correlation does not necessarily imply causation, hence we cannot draw any further conclusions about the relationship between the diversity of authors and the governance of SE conferences and journals without conducting further qualitative and quantitative studies.

\section{Discussion: Insights and Mitigation Strategies}
\label{sec:discussion}

This paper presents one of the first quantitative studies that examines the diversity of the authorship and governance of three top academic conferences and two top academic journals in SE. The results obtained from this study indicate a need for further research to look into the factors affecting the diversity in the authorship and organization of SE conferences and journals. Moreover, such a study could also help analyze the effectiveness of these policy measures and aid in the creation of new Equity, Diversity, and Inclusion (EDI) policies.

\textbf{Insights.} One of the major findings is that the authors published at the top publication venues for Software Engineering have a low representation of people from Black and Hispanic ethnicities as compared to other ethnicities. While the ethnic diversity of the authorship at these venues trends upwards through the years for the conferences, the journals do not show any such clear trend. The representation of publications from South American and African countries are also comparatively low. However, there is no evidence of ethnic, gender, or location biases involved in citing a research paper accepted by these top conferences/journals. 

Another major insight from our results is that the PC, OC, and EB for all the conferences and journals are composed of more White members and more male members in comparison to non-White members and female members. Broadening the scope of EDI policies to accommodate and foster ethnic diversity, in addition to gender-based diversity, could be a potential strategy to improve the overall equity and diversity of these publication venues.

Another interesting observation from Figure~\ref{fig:rq1_geo} is that the percentage of papers authored from European institutions is higher among the two selected journals as compared to the three conferences, and even exceeds the percentage of papers published in North America and Asia. Further investigation may be required to see if this observation generalizes to other SE journals as well.

\textbf{Mitigation Strategies.} A potential strategy to improve the ethnic diversity in the top conferences is to diversify the location of the conferences to include countries from different parts of the world, such as Africa and South America, as opposed to countries from North America, Europe, or Oceania (which tend to be more White-dominated) and have been the locations of these conferences in the past. Hosting a conference in countries from underrepresented parts of the world could provide more opportunities for universities in those areas to showcase and publish their research in an internationally recognized venue, thus leading to an improvement in the resulting diversity of the authorship and organization of the conference. 

When designing EDI policies in SE conferences and journals, people belonging to the intersection of both a minority and a majority group may have a greater likelihood of being excluded from consideration in comparison to people belonging to the intersection of two or more minority communities ~\cite{asianWomenCS}. For instance, researchers have studied the intersectionality of women and minority ethnic groups, such as Black ethnicity, in SE ~\cite{blackWomenIntersectionality}. Authors and committee members belonging to these categories are doubly underrepresented due to being both women and belonging to a minority ethnic group. On the other hand, Asian women are underrepresented as they are women while simultaneously being overrepresented as Asians ~\cite{asianWomenCS}. Therefore, it is important to consider the challenges introduced by such intersectionalities when designing policies to improve EDI in SE conferences and journals.

\vspace{-0.2cm}
\section{Threats to Validity} 
\label{sec:threats}

In this section, we discuss the various threats to validity that may impact our methodology and conclusions, as well as how we address these threats.

\textbf{Internal Validity.}
We use Namsor ~\cite{namsor} to predict the ethnicity and gender of the authors and committee members using only their first and last names. To determine the accuracy of the predictions made by Namsor, we select a random sample of 400 names and manually determine their corresponding ethnicity and gender by searching each name on Google. We use the profile pictures from the personal websites, LinkedIn profile, or their author details on IEEE Xplore belonging to each name and determine the ethnicity and gender using our knowledge and judgement. We use the manually determined ethnicity and gender values as the ground truth and check the accuracy of the ethnicity and gender values as predicted by Namsor. We find that for our sample, Namsor has an accuracy of 85.25\% in predicting the ethnicity and 90.63\% in predicting the gender, in accordance with the ground truth. We acknowledge that our manual labeling process may introduce researcher bias and limit the generalizability of our findings. However, we believe that this method provides a reasonable estimate of Namsor’s performance and reliability.

Additionally, the tool used to extract the gender and ethnicity (according to the classifications used by the US census) using just the given name and surname of a person is NamSor, which is comparatively more accurate than other tools for gender detection ~\cite{namsor-gender-performance} as well as ethnicity detection ~\cite{namsor-eth-performance}, but is not 100\% accurate. This lack of total accuracy introduces a margin of error in our classification of people's ethnicity and gender.

Furthermore, the ethnicity of a person is not necessarily the same as their race according to the US census classification. We also recognize that using the US census classification for determining the perceived ethnicity of a given name is a US-centric perspective. Nevertheless, we choose the "Name US Race" feature on NamSor as a benchmark for the classification of a person's perceived ethnicity.

Moreover, our study aims to study the relationship between the perceived ethnicity of people and their participation in SE conferences and journals. While a person's ethnic identity may be more complex, we rely on the names of the people to predict their most likely ethnicity as would be perceived by reviewers and people citing their works.

\textbf{External Validity.}  
We focus our analysis to the top three conferences and two journals in SE. As these have a rigorous peer review process that ensures that only high quality research is published, we attempt to investigate whether these venues, in particular, allow a wider and more diverse group of researchers to publish at these venues. However, the conclusions we derive from our results may not generalize to all SE conferences and journals.

\textbf{Construct Validity.} 
Our study measures the diversity in conferences and journals but only includes the data from the papers accepted by the conferences and journals and excludes any data about papers that were submitted and not accepted by these venues. Some of these venues, in some years, engage in double-blind peer review to mitigate reviewer biases towards the authors based on their ethnicity, gender identity, or geographical location.

\textbf{Reliability.} 
The lack of availability of historical data on the EB members for IEEE TSE (from 2014 onwards) limits our analysis. We reached out to the administration of IEEE TSE via email to procure the missing data about the EB members. However, we did not receive any response. Thus, the findings and conclusions derived about the governance of IEEE TSE, in particular, are valid only for the years 2010-2013. This missing data limits the generalizability of the findings and conclusions pertaining to the diversity of the EB in SE academic journals. 





\section{Related Work}
\label{sec:relatedwork}

In their work, Perez et al. ~\cite{se-diversity} perform a systematic literature review of 131 previous diversity studies in the SE industry and OSS development, which proposes to increase the awareness and reduce the systematic biases towards minority communities. The authors focus on gender, age, race and nationality as the four dimensions of diversity and find the existence of a strong bias against women in the SE industry and OSS development. They show that race has been the least studied dimension of diversity in SE and requires further investigation. In our work, we quantitatively analyze the state of diversity in SE academic conferences and journals by considering authorship as well as governance of these venues.

For the purpose of providing actionable information about the state of SE research, Vasilescu et al. ~\cite{se-conf-dataset} curate a dataset from eleven SE academic conferences from the years 2004 to 2012. The dataset includes information about the authors and the PC members of the eleven SE conferences. The authors intend for the dataset to be used by the steering or program committees to assess the selection process of submitted papers, and for authors to decide where to publish. However, the dataset lacks the demographic data required to analyze the state of the diversity of SE research based on ethnicity, gender or geographical location of the authors and governance committee members. Our work extends the availability of this form of data about SE publication venues and analyzes the data through our research questions.

Bano and Zowghi ~\cite{governance_diversity} present a qualitative study examining the gender disparity in the roles of the general chair, program chair and the main track PC members in six highly ranked SE conferences over a period of ten years. Their study curates the opinions of ten participants who have served at some of these conferences in leadership roles. Bano and Zowghi find significantly lesser number of women in top roles in the selected SE conferences. They also observe that having a female general chair or PC chair is not always correlated to the increase in participation of women in other visible organizational roles. Their study relies on qualitative methods to gather data about the gender diversity in the organization of SE conferences. In contrast, our study sheds light on the current state of ethnicity, gender and geographical diversity of both authorship and governance using quantitative measures, while also investigating any correlation between them. Furthermore, we extend our quantitative analysis to include two major SE journals in addition to three top SE conferences to broaden the scope of our study and enhance the generalizability of our results across SE academia.

\section{Conclusion} 
\label{sec:conclusion}

The quantitative study performed in this paper investigates the ethnicity, gender, and geographical diversity in academic publications in three top academic conferences and two major journals in Software Engineering. 
The study focuses on three conferences - ICSE (main track), FSE, and ASE, and two journals - IEEE TSE and ACM TOSEM. We programmatically collected data about the authorship and governance from various sources (DBLP, SCOPUS, IEEE Xplore, ACM Digital Library). We use NamSor to determine the perceived ethnicity and gender of individuals. The study poses four research questions that investigate the diversity of SE academic publication venues from 2010 to 2022. The results clearly indicate that the number of authors and the PC/OC/EB members belonging to certain genders, ethnicities, and geographical areas is significantly lesser as compared to the other categories. The ethnic diversity shows a positive (increasing) trend over the past decade across all conferences, suggesting that any measures adopted by these conferences to improve the diversity of publication have had a net positive impact. However, gender and geographical diversity do not show such a positive trend. We do not find any significant correlation between the diversity of authorship and that of the governance. We also find no evidence of bias in how often a paper is cited, based on the ethnicity, gender or geographical location of the first author. In the future, we plan to extend our analysis to conferences and journals in subdisciplines of SE. With access to the appropriate data, we could also analyze the diversity of the submissions to the conferences and journals rather than just the papers that were accepted. To study the entry barriers and specific challenges faced by researchers belonging to minority communities, qualitative studies involving surveys and interviews could help collect more data that could shed light on developing mitigation strategies.

\begin{acks}
We thank Rebecca Hutchinson, the University of Waterloo's liaison librarian for Computer Science for her assistance in accessing SCOPUS.
\end{acks}

\bibliographystyle{ACM-Reference-Format}
\bibliography{references}


\begin{thebibliography}{45}


\ifx \showCODEN    \undefined \def \showCODEN     #1{\unskip}     \fi
\ifx \showDOI      \undefined \def \showDOI       #1{#1}\fi
\ifx \showISBNx    \undefined \def \showISBNx     #1{\unskip}     \fi
\ifx \showISBNxiii \undefined \def \showISBNxiii  #1{\unskip}     \fi
\ifx \showISSN     \undefined \def \showISSN      #1{\unskip}     \fi
\ifx \showLCCN     \undefined \def \showLCCN      #1{\unskip}     \fi
\ifx \shownote     \undefined \def \shownote      #1{#1}          \fi
\ifx \showarticletitle \undefined \def \showarticletitle #1{#1}   \fi
\ifx \showURL      \undefined \def \showURL       {\relax}        \fi
\providecommand\bibfield[2]{#2}
\providecommand\bibinfo[2]{#2}
\providecommand\natexlab[1]{#1}
\providecommand\showeprint[2][]{arXiv:#2}

\bibitem[acm(2023)]%
        {acmDigitalLibrary}
 \bibinfo{year}{2023}\natexlab{}.
\newblock \bibinfo{title}{ACM Digital Library}.
\newblock \bibinfo{howpublished}{\url{https://dl.acm.org/}}.
\newblock
\newblock
\shownote{[Accessed 01-Feb-2023]}.


\bibitem[ese(2023)]%
        {esecfse2023}
 \bibinfo{year}{2023}\natexlab{}.
\newblock \bibinfo{title}{ACM Joint European Software Engineering Conference
  and Symposium on the Foundations of Software Engineering}.
\newblock \bibinfo{howpublished}{\url{https://2023.esec-fse.org/}}.
\newblock
\newblock
\shownote{[Accessed 25-Jan-2023]}.


\bibitem[tos(2023)]%
        {tosem}
 \bibinfo{year}{2023}\natexlab{}.
\newblock \bibinfo{title}{ACM Transactions on Software Engineering and
  Methodology}.
\newblock \bibinfo{howpublished}{\url{https://dl.acm.org/journal/tosem}}.
\newblock
\newblock
\shownote{[Accessed 25-Jan-2023]}.


\bibitem[Con(2023)]%
        {ConferenceRank2023}
 \bibinfo{year}{2023}\natexlab{}.
\newblock \bibinfo{title}{Best Computer Science Journals for Software
  Engineering \& Programming}.
\newblock
  \bibinfo{howpublished}{\url{https://research.com/conference-rankings/computer-science/software-programming}}.
\newblock
\newblock
\shownote{[Accessed 23-Jan-2023]}.


\bibitem[Jou(2023)]%
        {JournalRank2023}
 \bibinfo{year}{2023}\natexlab{}.
\newblock \bibinfo{title}{Best Computer Science Journals for Software
  Engineering \& Programming}.
\newblock
  \bibinfo{howpublished}{\url{https://research.com/journals-rankings/computer-science/software-programming}}.
\newblock
\newblock
\shownote{[Accessed 23-Jan-2023]}.


\bibitem[pyp(2023)]%
        {pypiCrossrefapi}
 \bibinfo{year}{2023}\natexlab{}.
\newblock \bibinfo{title}{Crossref API}.
\newblock
  \bibinfo{howpublished}{\url{https://pypi.org/project/crossrefapi/1.0.3/}}.
\newblock
\newblock
\shownote{[Accessed 25-Jan-2023]}.


\bibitem[DBL(2023)]%
        {DBLP}
 \bibinfo{year}{2023}\natexlab{}.
\newblock \bibinfo{title}{DBLP Computer Science Bibliography}.
\newblock \bibinfo{howpublished}{\url{https://dblp.org/}}.
\newblock
\newblock
\shownote{[Accessed 23-Jan-2023]}.


\bibitem[geo(2023)]%
        {geopy}
 \bibinfo{year}{2023}\natexlab{}.
\newblock \bibinfo{title}{{G}eo{P}y 2.3.0}.
\newblock
  \bibinfo{howpublished}{\url{https://geopy.readthedocs.io/en/stable/}}.
\newblock
\newblock
\shownote{[Accessed 25-Jan-2023]}.


\bibitem[tse(2023)]%
        {tse}
 \bibinfo{year}{2023}\natexlab{}.
\newblock \bibinfo{title}{IEEE Transactions on Software Engineering}.
\newblock
  \bibinfo{howpublished}{\url{https://ieeexplore.ieee.org/xpl/aboutJournal.jsp?punumber=32}}.
\newblock
\newblock
\shownote{[Accessed 25-Jan-2023]}.


\bibitem[iee(2023)]%
        {ieeeIEEEXplore}
 \bibinfo{year}{2023}\natexlab{}.
\newblock \bibinfo{title}{IEEE Xplore}.
\newblock
  \bibinfo{howpublished}{\url{https://ieeexplore.ieee.org/Xplore/home.jsp}}.
\newblock
\newblock
\shownote{[Accessed 01-Feb-2023]}.


\bibitem[ase(2023)]%
        {ase2023}
 \bibinfo{year}{2023}\natexlab{}.
\newblock \bibinfo{title}{International Conference on Automated Software
  Engineering}.
\newblock
  \bibinfo{howpublished}{\url{https://conf.researchr.org/home/ase-2023}}.
\newblock
\newblock
\shownote{[Accessed 25-Jan-2023]}.


\bibitem[ics(2023)]%
        {icse}
 \bibinfo{year}{2023}\natexlab{}.
\newblock \bibinfo{title}{International Conference on Software Engineering}.
\newblock \bibinfo{howpublished}{\url{http://www.icse-conferences.org/}}.
\newblock
\newblock
\shownote{[Accessed 25-Jan-2023]}.


\bibitem[nam(2023)]%
        {namsor}
 \bibinfo{year}{2023}\natexlab{}.
\newblock \bibinfo{title}{{N}amsor: name checker for gender, origin and
  ethnicity classification}.
\newblock \bibinfo{howpublished}{\url{https://namsor.app/}}.
\newblock
\newblock
\shownote{[Accessed 25-Jan-2023]}.


\bibitem[pyb(2023)]%
        {pybliometrics}
 \bibinfo{year}{2023}\natexlab{}.
\newblock \bibinfo{title}{Pybliometrics: {P}ython-based {A}{P}{I}-{W}rapper to
  access {S}copus}.
\newblock
  \bibinfo{howpublished}{\url{https://pybliometrics.readthedocs.io/en/stable/}}.
\newblock
\newblock
\shownote{[Accessed 25-Jan-2023]}.


\bibitem[Baltes(2023)]%
        {dblp-retriever}
\bibfield{author}{\bibinfo{person}{Sebastian Baltes}.}
  \bibinfo{year}{2023}\natexlab{}.
\newblock \bibinfo{title}{GitHub - sbaltes/dblp-retriever: Retrieve paper
  metadata from conference proceedings and journals indexed in DBLP ---
  github.com}.
\newblock
  \bibinfo{howpublished}{\url{https://github.com/sbaltes/dblp-retriever}}.
\newblock
\newblock
\shownote{[Accessed 25-Jan-2023]}.


\bibitem[Baltes et~al\mbox{.}(2020)]%
        {social_factors_industry}
\bibfield{author}{\bibinfo{person}{Sebastian Baltes}, \bibinfo{person}{George
  Park}, {and} \bibinfo{person}{Alexander Serebrenik}.}
  \bibinfo{year}{2020}\natexlab{}.
\newblock \showarticletitle{Is 40 the new 60? how popular media portrays the
  employability of older software developers}.
\newblock \bibinfo{journal}{\emph{IEEE Software}} \bibinfo{volume}{37},
  \bibinfo{number}{6} (\bibinfo{year}{2020}), \bibinfo{pages}{26--31}.
\newblock


\bibitem[Bano and Zowghi(2019)]%
        {governance_diversity}
\bibfield{author}{\bibinfo{person}{Muneera Bano} {and} \bibinfo{person}{Didar
  Zowghi}.} \bibinfo{year}{2019}\natexlab{}.
\newblock \showarticletitle{Gender disparity in the governance of software
  engineering conferences}. In \bibinfo{booktitle}{\emph{2019 IEEE/ACM 2nd
  International Workshop on Gender Equality in Software Engineering (GE)}}.
  \bibinfo{pages}{21--24}.
\newblock


\bibitem[Blau(1977)]%
        {blau1977inequality}
\bibfield{author}{\bibinfo{person}{Peter~Michael Blau}.}
  \bibinfo{year}{1977}\natexlab{}.
\newblock \bibinfo{booktitle}{\emph{Inequality and heterogeneity: A primitive
  theory of social structure}}. Vol.~\bibinfo{volume}{7}.
\newblock


\bibitem[Chang(2023)]%
        {Chang2023}
\bibfield{author}{\bibinfo{person}{Kenneth Chang}.}
  \bibinfo{year}{2023}\natexlab{}.
\newblock \showarticletitle{Asian Researchers Face Disparity With Key U.S.
  Science Funding Source}.
\newblock \bibinfo{journal}{\emph{The New York Times}} (\bibinfo{date}{Jan}
  \bibinfo{year}{2023}).
\newblock
\urldef\tempurl%
\url{https://www.nytimes.com/2023/01/04/science/asian-scientists-nsf-funding.html}
\showURL{%
\tempurl}


\bibitem[Chisholm et~al\mbox{.}(1999)]%
        {MIT1999}
\bibfield{author}{\bibinfo{person}{Sallie~W. Chisholm} {et~al\mbox{.}}}
  \bibinfo{year}{1999}\natexlab{}.
\newblock \bibinfo{title}{MIT Faculty Newsletter: A study on the status of
  women faculty in science at MIT}.
\newblock
\newblock
\urldef\tempurl%
\url{http://web.mit.edu/fnl/women/women.html}
\showURL{%
\tempurl}


\bibitem[Cochran et~al\mbox{.}(2020)]%
        {blackWomenIntersectionality}
\bibfield{author}{\bibinfo{person}{Geraldine~L Cochran},
  \bibinfo{person}{Mildred Boveda}, {and} \bibinfo{person}{Chanda
  Prescod-Weinstein}.} \bibinfo{year}{2020}\natexlab{}.
\newblock \showarticletitle{Intersectionality in STEM education research}.
\newblock In \bibinfo{booktitle}{\emph{Handbook of research on STEM
  education}}. \bibinfo{pages}{257--266}.
\newblock


\bibitem[de~Souza and Gama(2020)]%
        {industry_diversity}
\bibfield{author}{\bibinfo{person}{Nat{\'a}lia Pinheiro~Ramos de Souza} {and}
  \bibinfo{person}{Kiev Gama}.} \bibinfo{year}{2020}\natexlab{}.
\newblock \showarticletitle{Diversity and Inclusion: Culture and Perception in
  Information Technology Companies}.
\newblock \bibinfo{journal}{\emph{IEEE Revista Iberoamericana de Tecnologias
  del Aprendizaje}} \bibinfo{volume}{15}, \bibinfo{number}{4}
  (\bibinfo{year}{2020}), \bibinfo{pages}{352--361}.
\newblock


\bibitem[Díaz-García et~al\mbox{.}(2013)]%
        {DiazGarcia2013}
\bibfield{author}{\bibinfo{person}{Cristina Díaz-García},
  \bibinfo{person}{Angela González-Moreno}, {and}
  \bibinfo{person}{Francisco~Jose Sáez-Martínez}.}
  \bibinfo{year}{2013}\natexlab{}.
\newblock \showarticletitle{Gender diversity within R\&D teams: Its impact on
  radicalness of innovation}.
\newblock \bibinfo{journal}{\emph{Innovation}} \bibinfo{volume}{15},
  \bibinfo{number}{2} (\bibinfo{year}{2013}), \bibinfo{pages}{149--160}.
\newblock
\urldef\tempurl%
\url{https://doi.org/10.5172/impp.2013.15.2.149}
\showDOI{\tempurl}


\bibitem[Elsevier(2023)]%
        {scopus}
\bibfield{author}{\bibinfo{person}{Elsevier}.} \bibinfo{year}{2023}\natexlab{}.
\newblock \bibinfo{title}{Scopus - Abstract and Citation Database}.
\newblock \bibinfo{howpublished}{\url{https://www.scopus.com/home}}.
\newblock
\newblock
\shownote{[Accessed 25-Jan-2023]}.


\bibitem[Ford et~al\mbox{.}(2019)]%
        {social_factors_oss}
\bibfield{author}{\bibinfo{person}{Denae Ford}, \bibinfo{person}{Mahnaz
  Behroozi}, \bibinfo{person}{Alexander Serebrenik}, {and}
  \bibinfo{person}{Chris Parnin}.} \bibinfo{year}{2019}\natexlab{}.
\newblock \showarticletitle{Beyond the code itself: how programmers really look
  at pull requests}. In \bibinfo{booktitle}{\emph{2019 IEEE/ACM 41st
  International Conference on Software Engineering: Software Engineering in
  Society)}}. \bibinfo{pages}{51--60}.
\newblock


\bibitem[Frachtenberg and Koster(2020)]%
        {compsysauthors}
\bibfield{author}{\bibinfo{person}{Eitan Frachtenberg} {and}
  \bibinfo{person}{Noah Koster}.} \bibinfo{year}{2020}\natexlab{}.
\newblock \showarticletitle{A survey of accepted authors in computer systems
  conferences}.
\newblock \bibinfo{journal}{\emph{PeerJ Computer Science}}  \bibinfo{volume}{6}
  (\bibinfo{date}{Sept.} \bibinfo{year}{2020}), \bibinfo{pages}{e299}.
\newblock
\showISSN{2376-5992}
\urldef\tempurl%
\url{https://doi.org/10.7717/peerj-cs.299}
\showDOI{\tempurl}


\bibitem[Freire et~al\mbox{.}(2021)]%
        {aiconf}
\bibfield{author}{\bibinfo{person}{Ana Freire}, \bibinfo{person}{Lorenzo
  Porcaro}, {and} \bibinfo{person}{Emilia Gómez}.}
  \bibinfo{year}{2021}\natexlab{}.
\newblock \showarticletitle{Measuring Diversity of Artificial Intelligence
  Conferences}. In \bibinfo{booktitle}{\emph{AAAI Workshop on Diversity in
  Artificial Intelligence 2021}}. \bibinfo{pages}{39--50}.
\newblock


\bibitem[Girden(1992)]%
        {anova-test}
\bibfield{author}{\bibinfo{person}{Ellen~R Girden}.}
  \bibinfo{year}{1992}\natexlab{}.
\newblock \bibinfo{booktitle}{\emph{ANOVA: Repeated measures}}.
\newblock \bibinfo{publisher}{Sage}.
\newblock
\urldef\tempurl%
\url{https://doi.org/10.4135/9781412983419}
\showDOI{\tempurl}


\bibitem[Guizani et~al\mbox{.}(2022)]%
        {edi-oss}
\bibfield{author}{\bibinfo{person}{Mariam Guizani}, \bibinfo{person}{Bianca
  Trinkenreich}, \bibinfo{person}{Aileen~Abril Castro-Guzman},
  \bibinfo{person}{Igor Steinmacher}, \bibinfo{person}{Marco Gerosa}, {and}
  \bibinfo{person}{Anita Sarma}.} \bibinfo{year}{2022}\natexlab{}.
\newblock \showarticletitle{Perceptions of the State of D\&I and D\&I
  Initiative in the ASF}. In \bibinfo{booktitle}{\emph{Proceedings of the 2022
  ACM/IEEE 44th International Conference on Software Engineering: Software
  Engineering in Society}}. \bibinfo{pages}{130–142}.
\newblock
\showISBNx{9781450392273}
\urldef\tempurl%
\url{https://doi.org/10.1145/3510458.3513008}
\showDOI{\tempurl}


\bibitem[Hendricks(2023)]%
        {crossref}
\bibfield{author}{\bibinfo{person}{Ginny Hendricks}.}
  \bibinfo{year}{2023}\natexlab{}.
\newblock \bibinfo{title}{Crossref}.
\newblock \bibinfo{howpublished}{\url{https://www.crossref.org/}}.
\newblock
\newblock
\shownote{[Accessed 02-Feb-2023]}.


\bibitem[Joecks et~al\mbox{.}(2013)]%
        {Joecks2013}
\bibfield{author}{\bibinfo{person}{Jasmin Joecks}, \bibinfo{person}{Kerstin
  Pull}, {and} \bibinfo{person}{Karin Vetter}.}
  \bibinfo{year}{2013}\natexlab{}.
\newblock \showarticletitle{Gender Diversity in the Boardroom and Firm
  Performance: What Exactly Constitutes a “Critical Mass?”}.
\newblock \bibinfo{journal}{\emph{Journal of Business Ethics}}
  \bibinfo{volume}{118}, \bibinfo{number}{1} (\bibinfo{year}{2013}),
  \bibinfo{pages}{61--72}.
\newblock
\showISSN{1573-0697}
\urldef\tempurl%
\url{https://doi.org/10.1007/s10551-012-1553-6}
\showDOI{\tempurl}


\bibitem[Lin(2021)]%
        {asianWomenCS}
\bibfield{author}{\bibinfo{person}{Stephanie~C Lin}.}
  \bibinfo{year}{2021}\natexlab{}.
\newblock \emph{\bibinfo{title}{“The Modern Mulan of Tech”: A Critical
  Analysis of Asian American Women and their Experiences in Computer Science}}.
\newblock \bibinfo{thesistype}{Ph.\,D. Dissertation}.
  \bibinfo{school}{University of Illinois Urbana-Champaign}.
\newblock


\bibitem[Nelson(2014)]%
        {Nelson2014}
\bibfield{author}{\bibinfo{person}{Beryl Nelson}.}
  \bibinfo{year}{2014}\natexlab{}.
\newblock \showarticletitle{{The data on diversity}}.
\newblock \bibinfo{journal}{\emph{Commun. ACM}} \bibinfo{volume}{57},
  \bibinfo{number}{11} (\bibinfo{year}{2014}), \bibinfo{pages}{86--95}.
\newblock
\showISSN{00010782}
\urldef\tempurl%
\url{https://doi.org/10.1145/2597886}
\showDOI{\tempurl}


\bibitem[{\O}stergaard et~al\mbox{.}(2011)]%
        {se-productivity}
\bibfield{author}{\bibinfo{person}{Christian~R {\O}stergaard},
  \bibinfo{person}{Bram Timmermans}, {and} \bibinfo{person}{Kari Kristinsson}.}
  \bibinfo{year}{2011}\natexlab{}.
\newblock \showarticletitle{Does a different view create something new? The
  effect of employee diversity on innovation}.
\newblock \bibinfo{journal}{\emph{Research policy}} \bibinfo{volume}{40},
  \bibinfo{number}{3} (\bibinfo{year}{2011}), \bibinfo{pages}{500--509}.
\newblock


\bibitem[Rodr\'{\i}guez-P\'{e}rez et~al\mbox{.}(2021)]%
        {se-diversity}
\bibfield{author}{\bibinfo{person}{Gema Rodr\'{\i}guez-P\'{e}rez},
  \bibinfo{person}{Reza Nadri}, {and} \bibinfo{person}{Meiyappan Nagappan}.}
  \bibinfo{year}{2021}\natexlab{}.
\newblock \showarticletitle{Perceived Diversity in Software Engineering: A
  Systematic Literature Review}.
\newblock \bibinfo{journal}{\emph{Empirical Software Engineering}}
  \bibinfo{volume}{26}, \bibinfo{number}{5} (\bibinfo{date}{sep}
  \bibinfo{year}{2021}), \bibinfo{numpages}{38}~pages.
\newblock
\showISSN{1382-3256}
\urldef\tempurl%
\url{https://doi.org/10.1007/s10664-021-09992-2}
\showDOI{\tempurl}


\bibitem[Scott and Knott(1974)]%
        {scott-knott-test}
\bibfield{author}{\bibinfo{person}{Alastair Scott} {and}
  \bibinfo{person}{Martin Knott}.} \bibinfo{year}{1974}\natexlab{}.
\newblock \showarticletitle{A Cluster Analysis Method for Grouping Means in the
  Analysis of Variance}.
\newblock \bibinfo{journal}{\emph{Biometrics}}  \bibinfo{volume}{30}
  (\bibinfo{year}{1974}), \bibinfo{pages}{507}.
\newblock


\bibitem[Sebo(2021)]%
        {namsor-gender-performance}
\bibfield{author}{\bibinfo{person}{Paul Sebo}.}
  \bibinfo{year}{2021}\natexlab{}.
\newblock \showarticletitle{Performance of gender detection tools: a
  comparative study of name-to-gender inference services}.
\newblock \bibinfo{journal}{\emph{Journal of the Medical Library Association:
  JMLA}} \bibinfo{volume}{109}, \bibinfo{number}{3} (\bibinfo{year}{2021}),
  \bibinfo{pages}{414}.
\newblock


\bibitem[Sebo(2022)]%
        {namsor-eth-performance}
\bibfield{author}{\bibinfo{person}{Paul Sebo}.}
  \bibinfo{year}{2022}\natexlab{}.
\newblock \bibinfo{title}{NamSor's performance in predicting the country of
  origin and ethnicity of 90,000 researchers based on their first and last
  names}.
\newblock
\newblock
\urldef\tempurl%
\url{https://doi.org/10.21203/rs.3.rs-1565759/v3}
\showDOI{\tempurl}
\newblock
\shownote{Preprint on webpage at
  \url{https://www.researchsquare.com/article/rs-1565759/v3}}.


\bibitem[Serebrenik et~al\mbox{.}(2020)]%
        {policy_diversity}
\bibfield{author}{\bibinfo{person}{Alexander Serebrenik},
  \bibinfo{person}{Kelly Blincoe}, \bibinfo{person}{Byron Williams}, {and}
  \bibinfo{person}{Joanne Atlee}.} \bibinfo{year}{2020}\natexlab{}.
\newblock \showarticletitle{Diversity and Inclusion in the Software Engineering
  Research Community}.
\newblock \bibinfo{journal}{\emph{ACM Special Interest Group on Software
  Engineering}} \bibinfo{volume}{45}, \bibinfo{number}{4}
  (\bibinfo{year}{2020}), \bibinfo{pages}{5--7}.
\newblock


\bibitem[Tourani et~al\mbox{.}(2017)]%
        {se-productivity-2}
\bibfield{author}{\bibinfo{person}{Parastou Tourani}, \bibinfo{person}{Bram
  Adams}, {and} \bibinfo{person}{Alexander Serebrenik}.}
  \bibinfo{year}{2017}\natexlab{}.
\newblock \showarticletitle{Code of conduct in open source projects}.
  \bibinfo{pages}{24--33}.
\newblock
\urldef\tempurl%
\url{https://doi.org/10.1109/SANER.2017.7884606}
\showDOI{\tempurl}


\bibitem[Vasilescu et~al\mbox{.}(2015a)]%
        {oss_diversity}
\bibfield{author}{\bibinfo{person}{Bogdan Vasilescu}, \bibinfo{person}{Vladimir
  Filkov}, {and} \bibinfo{person}{Alexander Serebrenik}.}
  \bibinfo{year}{2015}\natexlab{a}.
\newblock \showarticletitle{Perceptions of diversity on git hub: A user
  survey}. In \bibinfo{booktitle}{\emph{2015 IEEE/ACM 8th International
  Workshop on Cooperative and Human Aspects of Software Engineering}}.
  \bibinfo{pages}{50--56}.
\newblock


\bibitem[Vasilescu et~al\mbox{.}(2015b)]%
        {se-productivity-3}
\bibfield{author}{\bibinfo{person}{Bogdan Vasilescu}, \bibinfo{person}{Daryl
  Posnett}, \bibinfo{person}{Baishakhi Ray}, \bibinfo{person}{Mark~G.J. van~den
  Brand}, \bibinfo{person}{Alexander Serebrenik}, \bibinfo{person}{Premkumar
  Devanbu}, {and} \bibinfo{person}{Vladimir Filkov}.}
  \bibinfo{year}{2015}\natexlab{b}.
\newblock \showarticletitle{Gender and Tenure Diversity in GitHub Teams}. In
  \bibinfo{booktitle}{\emph{Proceedings of the 33rd Annual ACM Conference on
  Human Factors in Computing Systems}}. \bibinfo{pages}{3789–3798}.
\newblock
\showISBNx{9781450331456}
\urldef\tempurl%
\url{https://doi.org/10.1145/2702123.2702549}
\showDOI{\tempurl}


\bibitem[Vasilescu et~al\mbox{.}(2013)]%
        {se-conf-dataset}
\bibfield{author}{\bibinfo{person}{Bogdan Vasilescu},
  \bibinfo{person}{Alexander Serebrenik}, {and} \bibinfo{person}{Tom Mens}.}
  \bibinfo{year}{2013}\natexlab{}.
\newblock \showarticletitle{A historical dataset of software engineering
  conferences}. In \bibinfo{booktitle}{\emph{2013 10th Working Conference on
  Mining Software Repositories (MSR)}}. \bibinfo{pages}{373--376}.
\newblock
\urldef\tempurl%
\url{https://doi.org/10.1109/MSR.2013.6624051}
\showDOI{\tempurl}


\bibitem[Woolley and Malone(2011)]%
        {Woolley2011}
\bibfield{author}{\bibinfo{person}{Anita Woolley} {and}
  \bibinfo{person}{Thomas~W. Malone}.} \bibinfo{year}{2011}\natexlab{}.
\newblock \showarticletitle{What Makes a Team Smarter? More Women}.
\newblock \bibinfo{journal}{\emph{Harvard Business Review}} (\bibinfo{date}{11}
  \bibinfo{year}{2011}).
\newblock


\bibitem[Zweben and Bizot(2022)]%
        {taulbee2021}
\bibfield{author}{\bibinfo{person}{Stuart Zweben} {and} \bibinfo{person}{Betsy
  Bizot}.} \bibinfo{year}{2022}\natexlab{}.
\newblock \showarticletitle{2021 Taulbee Survey: CS Enrollment Grows at All
  Degree Levels, With Increased Gender Diversity}.
\newblock \bibinfo{journal}{\emph{Computing Research News}}
  \bibinfo{volume}{34}, \bibinfo{number}{5} (\bibinfo{date}{May}
  \bibinfo{year}{2022}).
\newblock


\end{thebibliography}

\appendix

\end{document}